\documentstyle[spie,epsfig,a4]{article} 

\def\subsectfont{\raggedright\normalsize\bf}

\title{Near-Infrared-Spectroscopy with Extremely Large Telescopes:
Integral-Field- versus Multi-Object-Instruments} 

\author{F. Eisenhauer, M. Tecza, N. Thatte, S. Mengel, R. Hofmann
and R. Genzel
\skiplinehalf 
Max-Planck-Institut f\"ur extraterrestrische Physik, P.O.Box 1603, 85740 Garching, Germany
}

\begin{document} 
\maketitle 

\begin{abstract}
Integral-field-spectroscopy and multi-object-spectroscopy provide
the high multiplex gain required for efficient use of the upcoming
generation of extremely large telescopes. We present instrument 
developments and designs for both concepts, and how these designs
can be applied to cryogenic near-infrared instrumentation.
Specifically, the fiber-based concept stands out the possibility to
expand it to any number of image points, and its modularity predestines 
it to become the new concept for multi-field-spectroscopy.
Which of the three concepts --- integral-field-, multi-object-, or
multi-field-spectroscopy --- is best suited for the largest telescopes
is discussed considering the size of the objects and their 
density on the sky.

\end{abstract}


\section{Extremely Large Telescopes and Near-Infrared-Spectroscopy}

\subsection{Science with Extremely Large Telescopes}

Extremely Large Telescopes \cite{bash97,gilmozzi98,mountain97,sebring98} 
of the next century will have diameters of up to 100 m. 
Compared to current state of the art 10 m-class telescopes, 
the biggest of those telescopes provide a collecting power roughly 100 
times as big, and an angular resolution 10 times as good. The science-drivers
for such telescopes are threefold:

First, and most straight forward, we will be able to carry out spectroscopy 
of objects that we already know about from deep imaging, but which are too 
faint for spectroscopy with present day telescopes. The most prominent
and cited target of such observations is the Hubble Deep Field.

Second, we will image objects that we have never seen before, because they
are too faint or too distant. 

These two science drivers are the straight forward extension of what 
astronomers have done during the last century, and at first sight seem
related only
to the collecting area of the telescopes. But since such observations
will be background limited, we will only gain a factor of 10 --- 2.5 
magnitudes --- compared to the existing 10 m telescopes 
for seeing limited observations of point sources.
Only adaptive optics assisted observations at the diffraction limit 
of the telescopes will boost the limiting magnitude by a 
factor of 100 --- 5 magnitudes --- when enlarging the mirror-size
from 10 m to 100 m.
High angular resolution capability therefore will be mandatory.

And third, most exciting for us, is the prospect of exploring 
the universe at angular scales of a few milliarcseconds, the diffraction 
limit of such an Extremely Large Telescope. Like the Hubble Deep Field for
the faint object science, the direct imaging and spectroscopy 
of planets can serve as the final goal for high angular resolution astronomy. 
While imaging at this angular resolution will also be possible with 
interferometric arrays like the VLT, only the collecting area of 
several 1000 m$^2$ will provide enough photons for spectroscopy.

\subsection{The Need for Integral-Field- and Multi-Object-Instruments}

Since an Extremely Large Telescope will cost the order
of 1 billion US\$, throughput of the instruments has highest priority. 
Throughput in this context does not only mean 
imperfect transmission and detection of the light, but specifically
multiplex-gain. For the faint-object-science, best throughput implies 
simultaneous spectroscopy of as many objects as possible. This is
the standard domain of multi-object-spectroscopy. On the other hand,
if objects are to be resolved, and if we are interested in their 
complex structure, integral-field-spectroscopy is the the first choice.
This technology is definitely required when observing with adaptive
optics at the diffraction limit of a telescope, both to avoid
imperfect slit-positioning on the object, and for post-observational
correction of the imperfect point-spread-function by means of deconvolution.

\subsection{Why Near-Infrared-Spectroscopy?}

There are several reasons, both object-inherent and technical,
to carry out a significant fraction of the observations 
with such a telescope at near-infrared wavelengths:

First, many of the faint objects we are looking for --- like in the
Hubble Deep Field --- are at high redshift. 
Therefore a lot of the well established "optical" spectral 
diagnostics are shifted beyond 1 micron.

Second, many of the interesting objects in the universe --- like
nuclei of galaxies, star- and planet-forming regions --- are hidden 
behind dust. For example our Galactic Center is dimmed in the 
visible by about 30 magnitudes, while we suffer from only 
3 magnitudes of extinction in K-Band (2.2 $\mu$m).

And third, high angular resolution through the earth's
atmosphere is much easier achieved at longer wavelengths.
Even though there is no principle limit to achieve the diffraction limit 
in the visible, the high complexity of an adaptive optics system for an
extremely large telescopes with roughly $10^5$ actuators
\cite{gilmozzi98} may suggest to start with the easier task
of correcting in the near-infrared.

\section{Concepts for Integral-Field- and Multi-Object-Spectroscopy}

In this section we will present current developments and concepts 
for integral-field-spectroscopy and multi-object-spectroscopy 
(with emphasis on the technology developed at the Max-Planck-Institut
f\"ur extraterrestrische Physik), and outline the technology-challenge 
for their operation at cryogenic temperatures.

A number of instruments have been built or are going to be built for 
integral-field-spectroscopy and multi-object-spectroscopy.
Even though most of them are designed for operation at visible wavelengths,
their concepts are applicable for the near-infrared as well.
In this section we will describe the basic idea behind the 
different approaches and compare their specific properties and 
feasibility at cryogenic temperatures. 

\subsection{Integral-Field-Spectroscopy}

An integral-field-spectrograph obtains simultaneously the spectra 
for a two-dimensional field with a single exposure.
It therefore distinguishes itself from several other ways of measuring
spectra for a two-dimensional field, which all need multiple integrations.
Well known and applied in astronomy for several decades 
are (1) Fabry-Perot-imaging-spectroscopy, (2)
Fouriertransform-spectroscopy, (3) slit-scanning-spectroscopy. Why is
integral-field-spectroscopy most appropriate to ground-based
astronomy, especially at the highest angular resolution?

Fabry-Perot-imaging-spectroscopy and Fouriertransform-spectroscopy
require repetitive integrations to obtain full spectra. 
Therefore ground-based observations suffer a lot from varying atmospheric
conditions. Both atmospheric absorption and emission must be measured
in between two adjacent wavelength-settings, and since the 
atmospheric properties vary on a time-scale of minutes at 
near-infrared-wavelengths, long single exposures, and therefore high
quality spectra for faint objects are almost impossible to record with
wavelength scanning techniques.
The difference between integral-field-spectroscopy and slit-scanning is
in principle rather small, since both instruments provide roughly the same 
number of image points and the same spectral sampling. But because
most astronomical targets are far from being slit-like, 
almost all observations can gain a lot from a square field of view.

Three basic techniques are used in today's instruments for integral-field 
spectroscopy: The Mirror-Slicer, the Fiber-Slicer and the Micro-Pupil-Array. 

\begin{figure}[htp] 
\psfig{figure=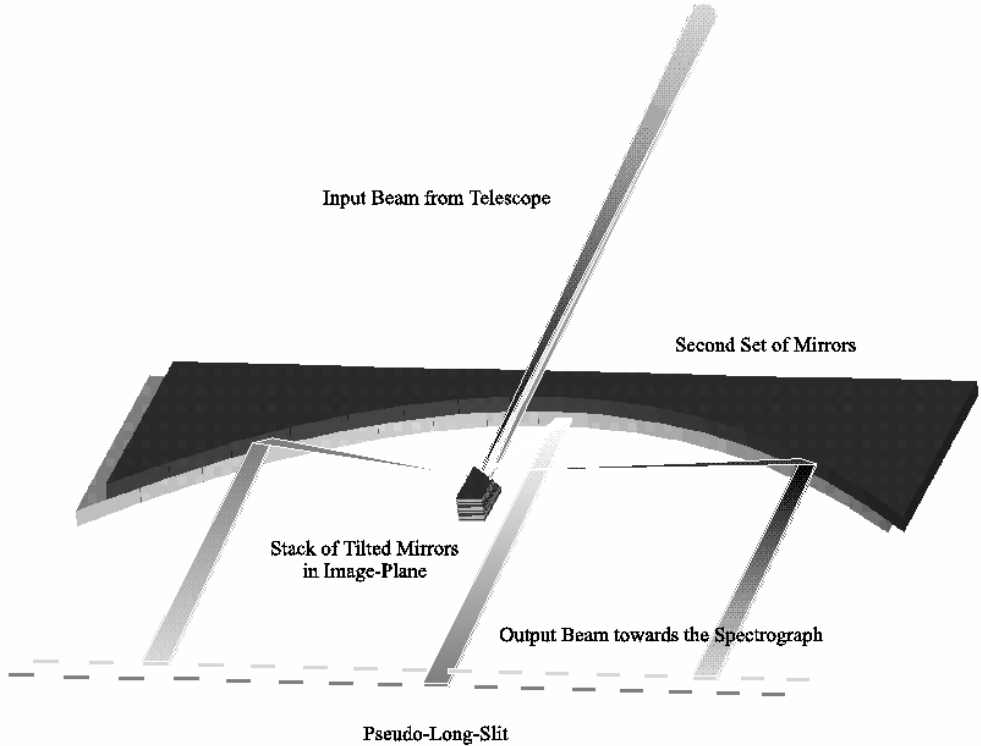,width=178mm}
\caption{\label{mirror}Image-slicing with mirrors:
A stack of tilted plane mirrors is placed in 
the focal plane, and each mirror reflects  the incoming light 
in a different direction. At a distance at which the rays 
from the different mirrors are clearly separated, a second set
of mirrors realigns the rays to form the long-slit.}
\end{figure}

The basic idea of image-slicing with mirrors is rather simple:
A stack of tilted plane mirrors is placed in 
the focal plane, and each mirror reflects the light from 
the image in a different direction. At a distance at which the rays 
from the different mirrors are clearly separated, a second set
of mirrors realigns the rays to form the long-slit (figure \ref{mirror})
of a long-slit-spectrograph, which disperses the light along the rows of 
the detector. This concept was successfully applied 
in the 3D-instrument \cite{krabbe97},
a near-infrared integral-field-spectrometer
developed and operated by MPE, and will be 
used for SPIFFI II, the adaptive-optics-assisted field-spectrometer for the 
VLT-instrument SINFONI \cite{thatte98,tecza98a} . 
The disadvantage of this concept is that shadowing 
at the steps of the first stack of mirrors leads to unavoidable light
losses. This shadowing effect increases with smaller mirrors and a larger
field of view. In order to have little light losses one 
would like to have large mirrors in the first stack. Because this 
increases the total slit-length, and therefore makes the
collimator of the spectrograph-optics uncomfortably big, 
a compromise 
has to be found. For SPIFFI II with its approximately 1000 spatial
pixels arranged  
in 32 slitlets, we chose the width of each mirror of the 
first stack  to be
300 $\mu$m, which leads to a total slit-length of about 300 mm.
For these parameters the shadowing-effect cuts out about 11\% of the
total light\cite{tecza99}. 
The whole slicer for SPIFFI II will be fabricated from 
Zerodur using classical polishing techniques.
Optical contacting of the individual mirrors will provide a monolithic
structure that is insensitive to changes in temperature.
With 3D we proved that the concept is feasible, and our recent
results from cool-downs of an engineering slicer to the temperature of liquid
nitrogen ensures operation in cryogenic instruments. 
However, the concept will find its limitation for 
much larger fields due to increased shadowing.
Other recent developments of mirror-slicers
derive from the basic design with plane mirrors, and take advantage 
of curved mirrors \cite{content97}. Such a concept avoids 
part of the shadowing-effects and provides a smaller 
"slit", thereby simplifying the design of the spectrograph-optics. 

A completely different approach for integral-field-spectroscopy is based
on optical fibers. In the image plane the two dimensional field is 
sampled by a bundle of optical fibers, which are then rearranged to 
a "long-slit". As in the mirror-slicer-concept, a normal long-slit
spectrograph can be used to disperse the light. As simple and expandable
as this concept seems, many little problems are inherent to such devices.

\begin{figure}[htp] 
\centering
\psfig{figure=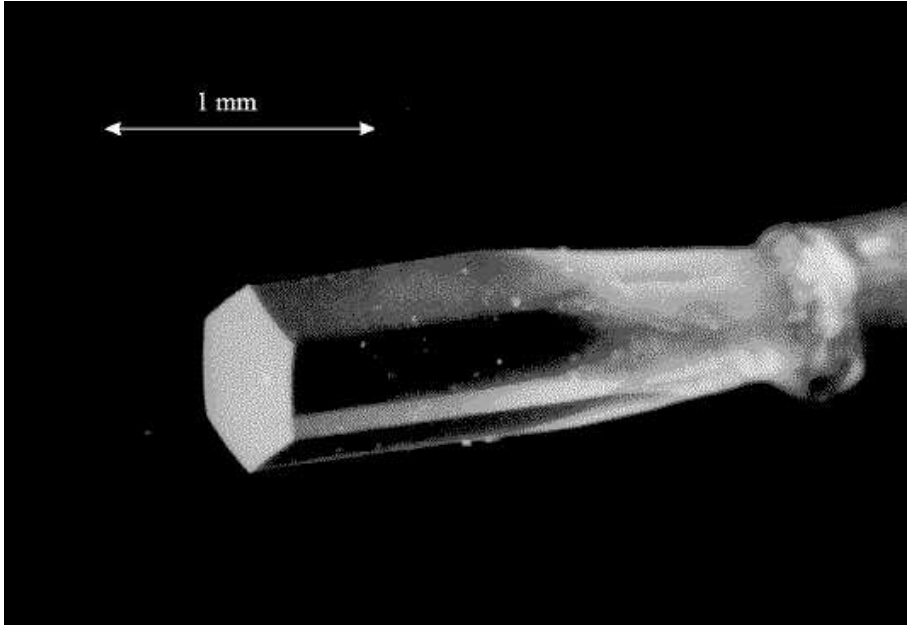,width=120mm}
\caption{\label{fiber}Flared fiber for cryogenic image-slicers:
The image shows one of the lens-fiber units that are used to sample the
image-plane. Each unit is built up from a silica-fiber, 
that has been  flared to the form of a taper, and a spherical lens
polished onto it.}
\end{figure}

To achieve a high coupling-efficiency, an array of square or hexagonal
lenslets with a filling factor of close to 100 \% is used to couple the
light into the fibers.  However, for a high coupling efficiency the
fibers have to be accurately positioned behind each lenslet and the image
quality of the lenslets has to be very good.  One way to loosen the
constraints on the positioning accuracy and the optical quality of the
lenslets is to use a larger $A \Omega$ for the fiber, which in turn
increases the f-number of the spectrograph-camera and finally limits
the pixel-size, especially at extremely large telescopes.
For a cryogenic instrument the positioning of the fibers behind the
lenslets is another critical point, since differential thermal
contraction both complicates gluing of the lenslets and fibers, and 
due to small displacements degrades the coupling efficiency.
For SPIFFI \cite{tecza98b} we therefore started the development of monolithic
lens-fiber-units (figure \ref{fiber}), each consisting of a 
silica-fiber, that has been flared 
and a spherical lens polished onto it \cite{tecza99}. Even though individual 
fiber-lens-units could be produced with an overall transmission of more 
than 70 \% --- including coupling efficiency, reflection losses 
and intrinsic absorption ---, the technology is not yet optimized
for producing several 1000 fibers at reasonable cost.
Despite all the technical problems with the image-slicer based
on flared fibers, there are three main advantages of this concept over
its competitors:  First, its possible extension to any number of fibers.
Second, this concept provides full flexibility in the arrangement of 
the fibers (see section on multiple-field-spectroscopy).
Third, the flared fiber is insensitive to a change in temperature and
can be used at cryogenic temperatures.
The flared-fiber technology will be implemented in LUCIFER \cite{mandel99}, 
the general-purpose near-infrared instrument for the LBT. 

Like the fiber-solution, the third concept 
for integral-field-spectroscopy is based on a micro-lens-array 
in the image plane.
But instead of reimaging the pupil of the 
individual lenses onto different fibers, the whole set of 
micro-pupils is now fed into a spectrograph \cite{bacon95}. With the
micro-pupils filling only a small fraction of the total field, 
with a slight tilt in the dispersion direction the spectra on the
detector fill the unused detector area between the micro-pupils without
overlapping of the individual spectra. Compared to the fiber-based concept, 
there is no additional loss of light due to the coupling of light into
the fiber. Also the technology of producing micro-lens-arrays
is now well established, and its application in cryogenic instruments 
seems straight forward. But while both mirror- and 
fiber-slicers can disperse the light all across the detector, the
spectra of the micro-pupils need to be truncated before they overlap
with the spectra of another micro-pupil. Therefore such instruments
can provide high spectral resolution only for a very limited 
wavelength range.

\subsection{Multi-Object-Spectroscopy}

The need for multi-object-spectroscopy is obvious  
for faint object astronomy. Whenever good statistics
is crucial for the scientific interpretation, we need to
have information on as many objects as possible. And since many 
programs require integration times of several hours per object,
simultaneous observations of a large number of objects are the
only possibility to carry out the observations within a
reasonable time.
 
There are two basic concepts to carry out simultaneous 
spectroscopy of multiple objects in a field: Using multiple slits
and coupling the light into fibers. 

In the multi-slit approach a mask with slits located at
the object positions is placed in the image-plane. This slit-plate
is normally fabricated "off-line" prior to the observation. 
Special care must be taken to avoid overlap of the spectra from
different slits. Therefore usually several masks and observations 
are required for a complete set of spectra of the objects within
a given field. The big advantage of such slit-mask-spectrographs is their
high optical throughput, since no additional optical element
is introduced.
Examples for such instruments are the CFHT MOS \cite{lefevre94}
and the two VIRMOS \cite{lefevre98} instruments for the VLT.
One way to overcome the "off-line-production" of the slit-masks
might be with micro-mirror arrays which would allow electronically
controlled object selection. 

The second concept of multi-object-spectroscopy is based on 
fibers. While in previous instruments fibers had to be placed by hand,
nowadays robots arrange the fibers, like in the AAT 2dF \cite{taylor97}. 
Depending on the image-scale and the f-number, the light is either
coupled directly into the fiber, or a lenslet is used to reimage 
the telescope-pupil onto the fiber core. As the fiber-based
integral-field-unit, 
such multi-object-spectrographs can be expanded to almost any 
number of objects. 

For the time being, no cryogenic multi-object-spectrograph
for the infrared wavelength range has been set into operation. For the 
LUCIFER instrument for the LBT, however, possible realization of the
two concepts --- multi-slit and fiber-based --- in a cryogenic 
instrument are under study: In a multi-slit-spectrograph, the 
technical key-problem is that the slit-masks have to be produced "off-line",
and need to be inserted into the cryogenic system. One possibility
would be an air-lock through which a set of slit-masks are fed into a
juke-box and cooled down to the temperature of liquid
nitrogen, before they are actually moved into the field.
A fiber-based system, however, will require a fully cryogenic 
robot to position the fibers. But unlike their optical counterparts,
present-day fibers for the infrared are either rather fragile
(zirconiumflouride), or show significant extinction towards longer
wavelengths (waterfree 
silica). Therefore long fibers and big movements should be avoided, 
and a "Spaltspinne"-like \cite{pitz93} mount of the fibers with a
long-slit-spectrograph located directly behind seems most
promising. While the fiber-technology --- e.g. based on the monolithic
concept described for the integral-field-unit --- is almost
established, a reliable cryogenic robot is not yet in operation.

A common problem to both multi-object-concepts described above is 
the need to have precise target positions. In addition no
(fiber-concept) or very limited (multi-slit, since all
slits are parallel) spatial information can be obtained.

\subsection{Multi-Field-Spectroscopy} 

Both problems of multi-object-spectrographs --- the need for precise 
target-positions, and the lack of spatial information --- will 
be overcome by multi-field-spectroscopy: Like in multi-object-spectrographs, 
multiple objects are observed simultaneously, but now each object 
is spatially sampled with an integral-field-unit.

\begin{figure}[htp] 
\centering
\psfig{figure=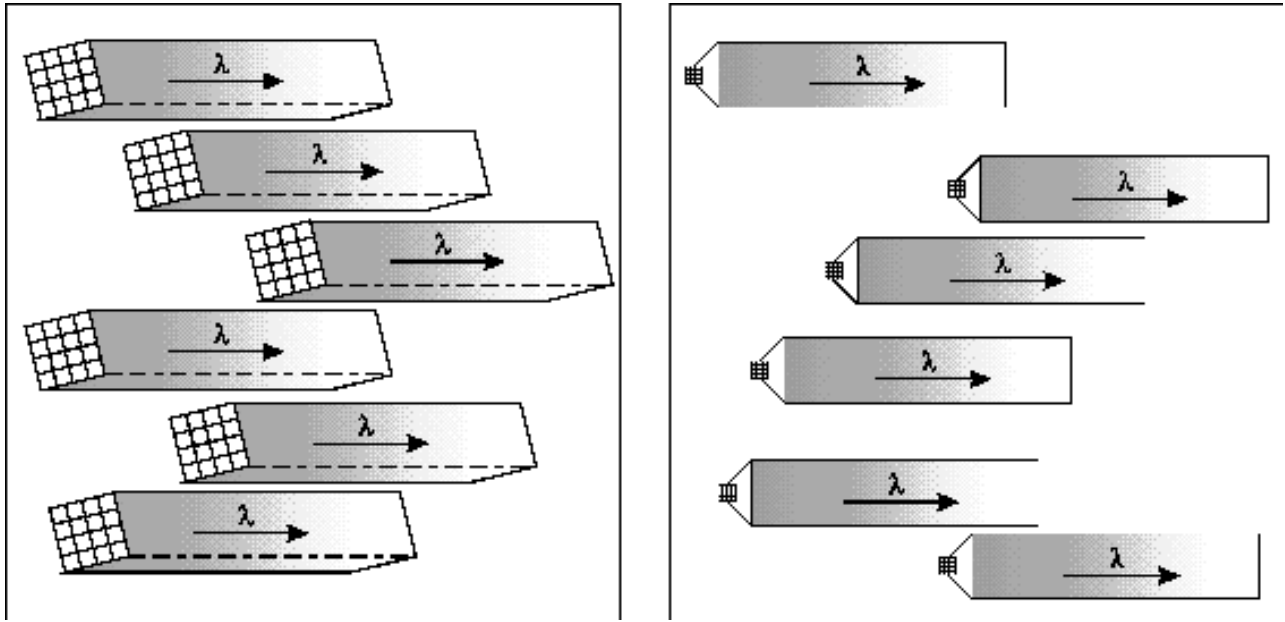,width=170mm}
\caption{\label{multifield}Mulit-field-spectroscopy with
micro-pupil-arrays (left) and multi-mirror-slicers (right) is the
natural extension of the slit-mask concept. Which slicer concept
matches best the science-program depends on the source density in the field. 
Please note that in both drawings the size of the individual fields
and pixels are the same. The relative separation of the individual
fields, however, is much larger for the mirror-slicer concept.  
}
\end{figure}

In principle each of the three basic concepts for
integral-field-units  --- mirror-slicer, fiber-slicer,
or micro-pupil-array --- could be combined with the multi-object
concept: 

Little mirror-slicers or micro-pupil-arrays
are the natural extension of the slit-mask (figure \ref{multifield}).
Assuming the same pixel-scale, the same size of the individual 
fields and the same size of the detector, 
the source density decides which slicer-concept matches best the
science-program. 
For micro-pupil-arrays, all the objects should be within a 
field with a linear dimension equal to the number of 
fields times the size of each single field. For mirror-slicers, the
objects should be arranged more loosely, the objects at least 
separated by the square of the linear dimension of each subfield.

\begin{figure}[htp] 
\centering
\psfig{figure=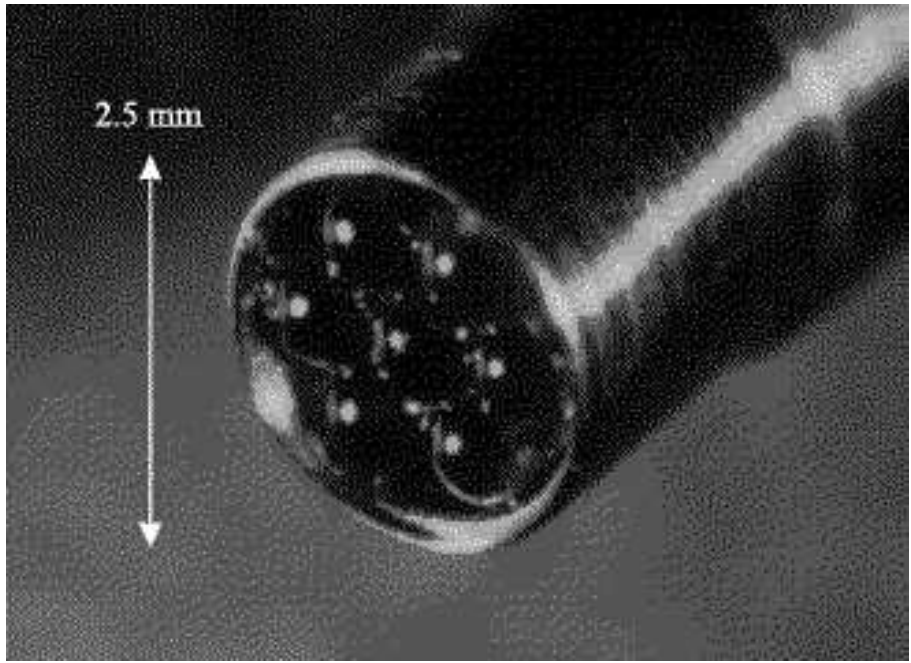,width=120mm}
\caption{\label{multifiber}Mulit-field-spectroscopy with fibers:
The monolithic lens-fiber-units of figure 2 can be arranged in any
form. This image shows the building block of a possible
multi-field-spectrograph that provides seven image points per object.}
\end{figure}

Most promising, however, is the combination of the fiber-based multi-object-
and integral-field-concepts. The single fibers of the
multi-object-spectrograph need only to be replaced by small fiber-slicers
built from several lenslet-fiber-units (figure
\ref{multifiber}). Depending on the 
science-program, either small individual fields with about seven 
pixels each, or larger fields with about 100 spatial elements would be
selected. The monolithic concept developed for the fiber-slicer 
as described above would fulfill all requirements for this kind
of multi-field-spectroscopy.

However, like multi-field-spectrographs based on fibers, all
multi-field-solutions for the infrared wavelength-range require
reliable cryogenic robots.

\section{Comparison}

\subsection{General Considerations for Extremely Large Telescopes}

Before comparing integral-field-, multi-object-, and
multi-field-spectroscopy for extremely large telescopes, we 
discuss the noise regime and pixel-scales for such instruments.

What is the maximum pixel scale? In order to get a rough estimate, we
will assume a telescope with a diameter of 100 m. The physical size
of a pixel of a present day near-infrared-detector is about
20 $\mu$m. We further know from present-day near-infrared-instruments 
like SPIFFI \cite{tecza98b} that the f-number of any camera-optic
needs to be greater than or equal to roughly 1 to achieve acceptable
image quality. From this limit, 
and the fact that $A\Omega$ is preserved in imaging optics, 
one can derive a maximum pixel size of 60 mas. 
So whenever larger image elements are
required, their flux must be spread over several pixels. 
The smallest pixel scale is determined by the diffraction limit of the
telescope. For the H-band (1.65 $\mu$m) the appropriate pixel scale to
Nyquist-sample the image is 3 mas.

What is the noise regime we have to work with? Let us assume H-band
observations again. Most of the sky-background in this wavelength
range arises from about 70 OH lines \cite{maihara93,oliva92}, 
which sum up to a total surface
brightness of about 14 mag / arcsec$^2$. The flux between the OH-lines
is roughly 18 mag / arcsec$^2$. The first lesson we
learn from these numbers is that even for present day technology 
OH-suppression is crucial for deep observations. In order to loose only
1/10 of the H-band-spectrum to OH-contaminated pixels, roughly
1400 pixels are required for Nyquist-sampling, corresponding to
a spectral resolution of approximately 3000 in H-band. But even at this 
spectral resolution and with adaptive-optics-pixel-scales, observations will be
background-limited assuming future detectors with a read-noise
close to 1 electron and negligible  dark-current, and
integration-times of the order of 1 hour.

\subsection{What concept is best suited for extremely large telescopes?}

Extremely large telescopes provide the
unprecedented opportunity for spectroscopy of (a) extremely small 
and (b) extremely faint objects.

For us it is obvious that spectroscopy at the diffraction
limit of an adaptive-optics equipped telescope and with pixel scales
of the order of milliarcseconds requires integral-field-units. 
Of the three concepts described above --- mirror-slicer,
fiber-slicer, micro-pupil-array ---, the mirror-slicer provides the most
efficient use of detector elements, because it is the only technology
that actually uses almost all pixels. Being not yet at its
limitation in field-coverage, this concept may be the choice for the next
generation of instruments. In the more distant future, when
detector-size and -availability will not limit our
instrumentation-plans any more, the micro-pupil-concept, and finally
the most expandable fiber-concept are most appropriate.

Since we will be sky-limited at near-infrared-wavelengths at any
pixel-scale, the biggest gain in sensitivity (5 magnitudes) 
over 10m-class telescopes will be achieved by adaptive-optics-assisted
observations  
of point-like sources. In order to make most efficient use of the 
telescope time, multi-object spectroscopy will be one of the most 
important operation modes for extremely large telescopes. But with the
pixel-size of a few milliarcseconds, the problem of accurate
slit-positioning will require the extension of the 
multi-object-technique towards the multi-field-approach.
For this application the fiber-based-concept combined with 
a cryogenic robot seems most promising.

For  spectroscopic surveys the object-density will finally 
determine the most appropriate instrumentation. But one should keep in
mind that --- in contrast to smaller but equally sensitive future
space-telescopes --- the maximum pixel-size for a 100m telescope will be
limited to about 50 mas. Assuming OH-suppressed observations in the
H-band with roughly 1400 spectral pixels for each image
point, even 16 detectors with 4k $\times$ 4k pixels could only cover
a field 20 arcsec on a side. In order to make efficient use of an 
integral-field-unit for this application, the source density should
be of the order $10^4$ objects per arcminute, like extragalactic 
star-forming regions. For most applications, however, the source
density will be much smaller, and the combination of deep imaging and
multi-field-spectroscopy will match best.

\end{document}